\documentclass{PoS}

\title{On the spectrum of QCD-like theories and the conformal window}

\ShortTitle{On the spectrum of QCD-like theories and the conformal window}

\author{Albert Deuzeman\\
Albert Einstein Center for Fundamental Physics - University of Bern, Switzerland\\
        E-mail: \email{a.deuzeman@itp.unibe.ch}}

\author{Maria Paola Lombardo\\
    INFN-Laboratori Nazionali di Frascati, I-00044, Frascati (RM), Italy\\
        E-mail: \email{mariapaola.lombardo@lnf.infn.it}}

\author{\speaker{Elisabetta Pallante}\\
        Centre for Theoretical Physics, University of Groningen, 9747 AG, Netherlands\\
        E-mail: \email{e.pallante@rug.nl}}

\abstract{ We report on the spectrum of the $SU(3)$ gauge theory with twelve  flavours in the fundamental representation of the gauge group. We isolate distinctive features of the hadronic phase - the one proper of QCD at zero temperature - and the so called conformal phase. The latter should emerge at sufficiently large $N_f$ and before the loss of asymptotic freedom. 
In particular, we analyse available lattice data for the spectrum of $N_f=12$ and include a comparison with results with $N_f=16$; the latter theory, predicted by the perturbative 
$\beta$-function to develop an IRFP and therefore be in the conformal phase, can serve as a paradigm for the study of theories in the conformal window. 
Our analysis suggests that the theory with twelve flavours is in the conformal window, possibly close to its lower boundary. }

\FullConference{ The XXIX International Symposium on Lattice Field Theory - Lattice 2011\\
July 10-16, 2011\\
Squaw Valley, Lake Tahoe, California}

\begin{document}

\section{Introduction}
\label{sec:intro}

\begin{figure}[h]
\begin{center}
\includegraphics[width=.45\textwidth]{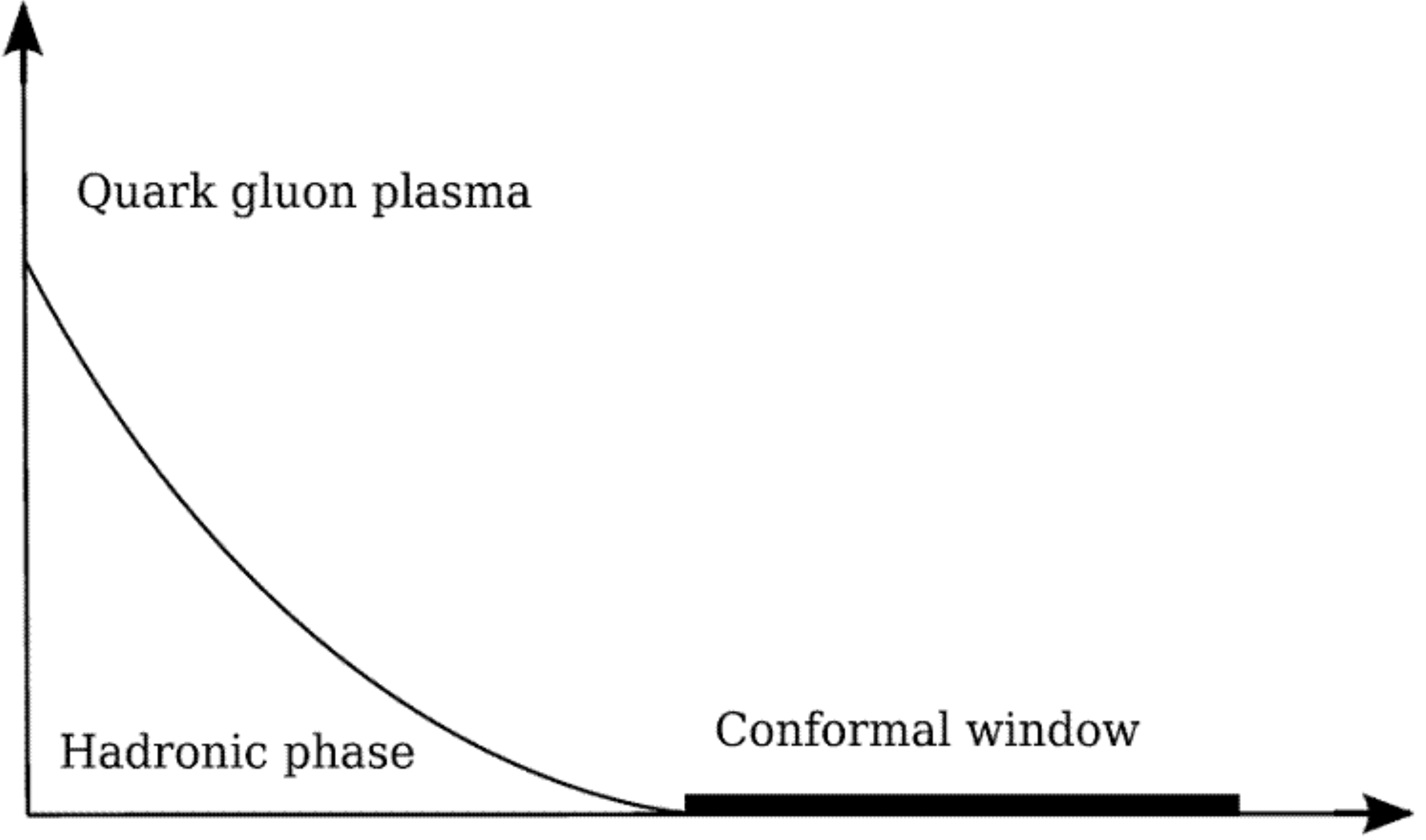}
\caption{The phase diagram of QCD-like theories in the $T, N_f$ plane.}
\label{fig:phase}
\end{center}
\end{figure}

The study in \cite{Deuzeman_PRD} for the theory with $N_f=12$ flavours supports a scenario as depicted in Fig.~\ref{fig:phase}, where the end-point of the chiral phase boundary signals the opening of the conformal window and $N_f=12$ is inside the conformal window. While conformality for the same theory has been reported by other groups \cite{Appelquist:2011dp,Hasenfratz:2011xn}, contrasting views were presented as well \cite{Fodor_Nf12}.
  
Any massless theory within the conformal window has exact chiral symmetry and develops an infrared fixed point (IRFP) at which the theory is conformal. 
Everywhere in the parameter space of the theory, except at the fixed point, observables will show only remnants of conformality. These remnants joint with the realization of exact chiral symmetry lead to features of the spectrum distinct from QCD.  In particular, adimensional mass ratios  are robust indicators of patterns of symmetries \cite{Deuzeman_PRD, Deuzeman:2010zz}.

The aim of our study, of which this proceeding is a preliminary account, is to isolate those features and use them in the comparison of $N_f=12$ lattice results with typical lattice QCD results. 
The set of data used here consists of the ensembles partly analysed in \cite{Deuzeman_PRD}, where more statistics has been collected for some points at the lattice bare couplings $\beta_L =6/g^2 =3.8,\,3.9$ and $4.0$, on the weak-coupling side of the bulk phase transition \cite{Deuzeman_PRD}. Data are at masses $am=0.07,0.06$ with volume $16^3$x$24$, $am=0.05$ with volume $24^4$ and $am=0.025$ with volume $32^4$. The action used is the tree-level Symanzik-improved gauge action with Asqtad staggered fermions. We also include the largest volume data reported in \cite{Fodor_Nf12} at one value of the bare lattice coupling $\beta_L =2.2$ and with a different improved lattice action; tree-level Symanzik improvement and two steps of stout-smearing in the staggered fermion matrix.
Whenever instructive, we compare with results for the theory with $N_f=16$ fundamental flavours from \cite{Damgaard:1997ut}. A study of $N_f=16$ with the same action as $N_f=12$ in \cite{Deuzeman_PRD} is in progress \cite{Nf16_us}. We defer to future work a more refined estimate of finite volume effects and a finite size scaling analysis.

In section \ref{sec:edinburgh} we analyse the  mass ratio $m_\pi /m_\rho$, the Edinburgh plot and additional ratios useful to discriminate between a QCD phase and a conformal phase. In section \ref{sec:spectrum} we consider another indicator of chiral symmetry restoration, the splitting between the vector and the axial ground states. Finally, in section \ref{sec:mpi_pbp} we analyse another key relation between the would be 
Goldstone boson and the chiral order parameter. We conclude in section \ref{sec:conclusions}.
%
%%%
\section{The Edinburgh plot and mass ratios.}
\label{sec:edinburgh}
One first significant spectrum observable is the ratio $m_\pi/m_\rho$, between the mass of the  
lightest pseudoscalar state (pion) $m_\pi$ and the mass of the lightest vector state (rho) $m_\rho$. In QCD at zero temperature, chiral symmetry is spontaneously broken and the pion is the (pseudo)Goldstone boson of the broken symmetry, implying that its mass will behave as $m_\pi\sim \sqrt{m}$. Instead, the vector mass contains a constant term and a leading correction linear in the quark mass, thus $m_\rho \sim m_0 +  b m$. In the chirally broken phase, modulo lattice artefacts,  one should thus expect their ratio to behave as $m_\pi/m_\rho\sim \sqrt{am}$ - as a function of the bare lattice quark mass  $am$. 

Within the conformal window chiral symmetry is restored.
The lightest pseudoscalar state is not anymore a Goldstone boson,
and there is no mass gap. 
At the IRFP and at infinite volume, the quark mass dependence of all hadron masses in the spectrum is governed by conformal symmetry: at leading order in the quark mass expansion all masses follow a power-law with common exponent determined by the anomalous dimension of the fermion mass operator at the IRFP. 
Hence we expect a constant ratio.  Away from the IRFP, for sufficiently light quarks and finite lattice volumes, the universal  power-law dependence receives corrections, due to the fact that the theory is interacting but no longer conformal.
\begin{figure}
\begin{minipage}[htb]{0.5\linewidth}
\begin{center}
\includegraphics[scale=0.25]{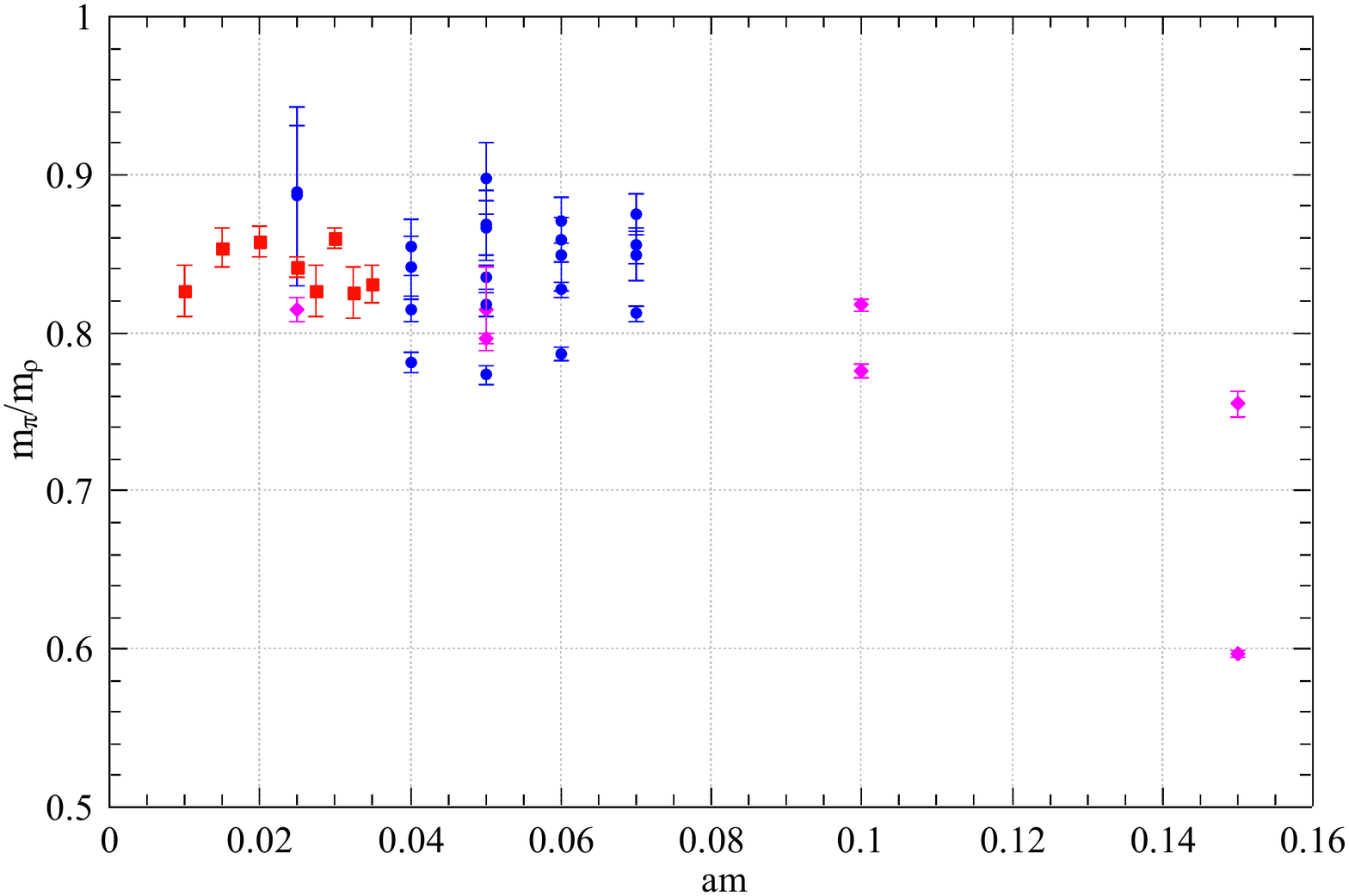}
\caption{Ratio $m_\pi/m_\rho$ as a function of the bare quark mass for all existing data for $N_f=12$, and $N_f=16$: $N_f=12$ data from \cite{Fodor_Nf12} (red squares), $N_f=12$ data from this work and $\beta_L=3.8,3.9,4.0$ (blue circles), $N_f=16$ data  from \cite{Damgaard:1997ut} (magenta diamonds).   }
\label{fig:mpimrhoRatio}
\end{center}
\end{minipage}
\hspace{0.2cm}
\begin{minipage}[htb]{0.5\linewidth}
\begin{center}
\includegraphics[scale=0.25]{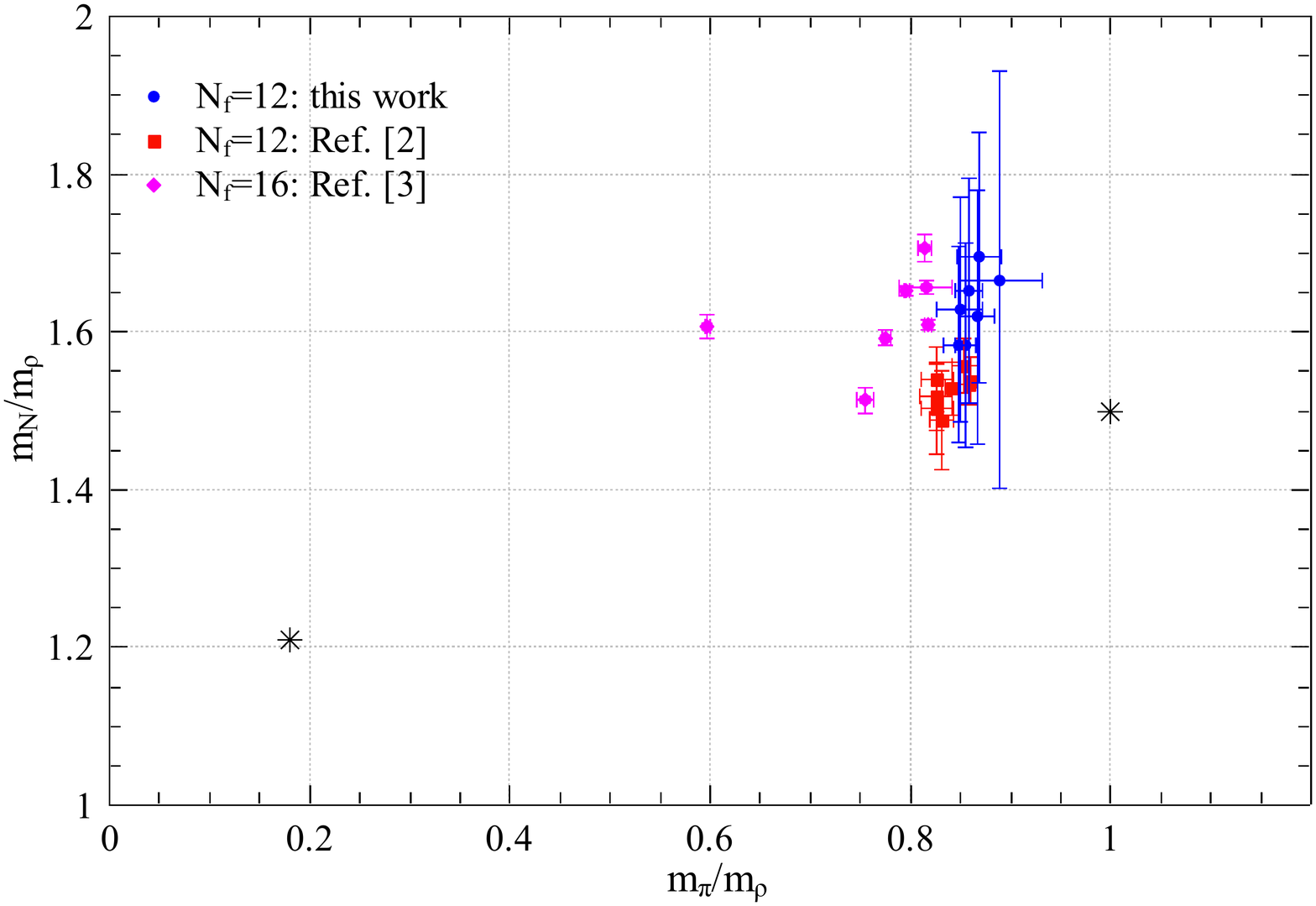}
\caption{Edinburgh plot: $N_f=12$ data from \cite{Fodor_Nf12} (red squares), $N_f=12$ data from this work and $\beta_L=3.8,3.9$ (blue circles), $N_f=16$ data from \cite{Damgaard:1997ut} (magenta diamonds). The QCD physical point (black star, leftmost) and the heavy quark limit (free theory) point (black star, rightmost) are shown.}
\label{fig:Edplot}
\end{center}
\end{minipage}
\end{figure}
Hence, the pseudoscalar-vector mass ratio is constant at the IRFP at infinite volume, and approximately constant in its surroundings and at finite volume, as it is the situation explored here.

In practice, the task remains the one of discriminating between a lattice mass ratio that goes to zero as $\sqrt{am}$ 
and a ratio that remains constant and O(1) 
over a significant range of masses. 

An important caveat concerns the extraction of mass eigenstates from correlators within the conformal window: correlators in the vicinity of the IRFP will follow a power law decay at leading order, corrected by mass contributions. However, for sufficiently large quark masses and away from the IRFP one expects lattice correlators to decay exponentially as in QCD, possibly with subleading conformal corrections. 
Given this caveat, we have analysed all correlators in this work assuming the standard multi-exponential time dependence. The pseudoscalar staggered correlator could be fitted with two cosine hyperbolic functions (fundamental and excited state) without the parity-odd (oscillating) component. All other staggered correlators could be fitted with a cosine hyperbolic with an oscillating component at intermediate and late times, with the addition of an excited state at early times. 
The largest uncertainties of the present analysis are related to the nucleon, for which more statistics and better smearing is needed. Temporal extents longer than $t=24,\, 32$ would also facilitate the analysis of correlators at the lightest masses. 

 Fig.~\ref{fig:mpimrhoRatio} shows that the mass ratio for all existing $N_f=12$ data is approximately constant over a wide range of  bare quark masses, as it should be  expected for a chirally symmetric theory.  Obviously, on the basis of these numerical evidences, we cannot exclude that a change of trend will occur at even lower masses.  Searching for combined evidences seems to be the optimal strategy for this case, following the line adopted in \cite{Deuzeman_PRD}. 

The Edinburgh plot, widely used in lattice QCD studies, is constructed in terms of adimensional ratios of masses and offers a powerful way to combine results of lattice calculations performed at different lattice spacings. Fig.~\ref{fig:Edplot} shows the Edinburgh plot for all existing data with $N_f=12$. For an instructive comparison, we also show lattice results for the $N_f=16$ theory from \cite{Damgaard:1997ut}; the latter is already known to be in the conformal window by perturbative arguments. 
The physical point of QCD (leftmost side of figure) corresponds to $m_\pi/m_\rho\simeq 0.18$ and $m_N/m_\rho\simeq 1.21$. On the other side of the figure,  a useful theoretical limit is the heavy quark mass limit (rightmost side of figure), where all masses in the spectrum are given by the sum of their valence quark masses, so that $m_\pi/m_\rho =1$ and $m_N/m_\rho =3/2$. This limit is also equivalent to the free theory limit. 
\begin{figure}
\begin{minipage}[t]{0.5\linewidth}
\begin{center}
\includegraphics[scale=0.27]{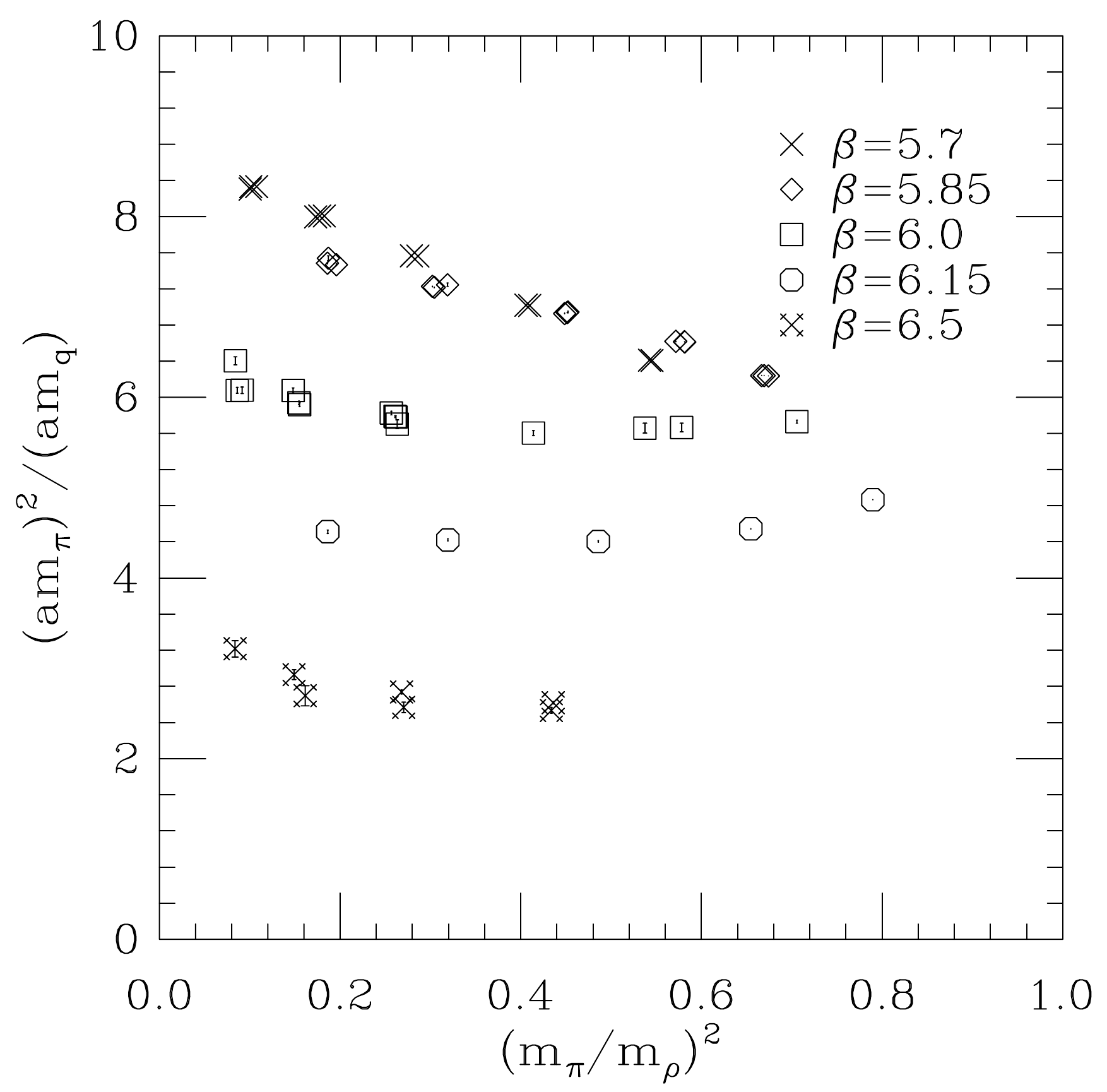}
\caption{Ratio $(am_\pi)^2/(am)$ as a function of $(m_\pi/m_\rho)^2$ for QCD data (quenched staggered) reported by the MILC collaboration \cite{MILC}. Notice that quenching should not affect the leading behaviour.  }
\label{fig:QCDlike}
\end{center}
\end{minipage}
\hspace{0.2cm}
\begin{minipage}[t]{0.5\linewidth}
\begin{center}
\includegraphics[scale=0.23]{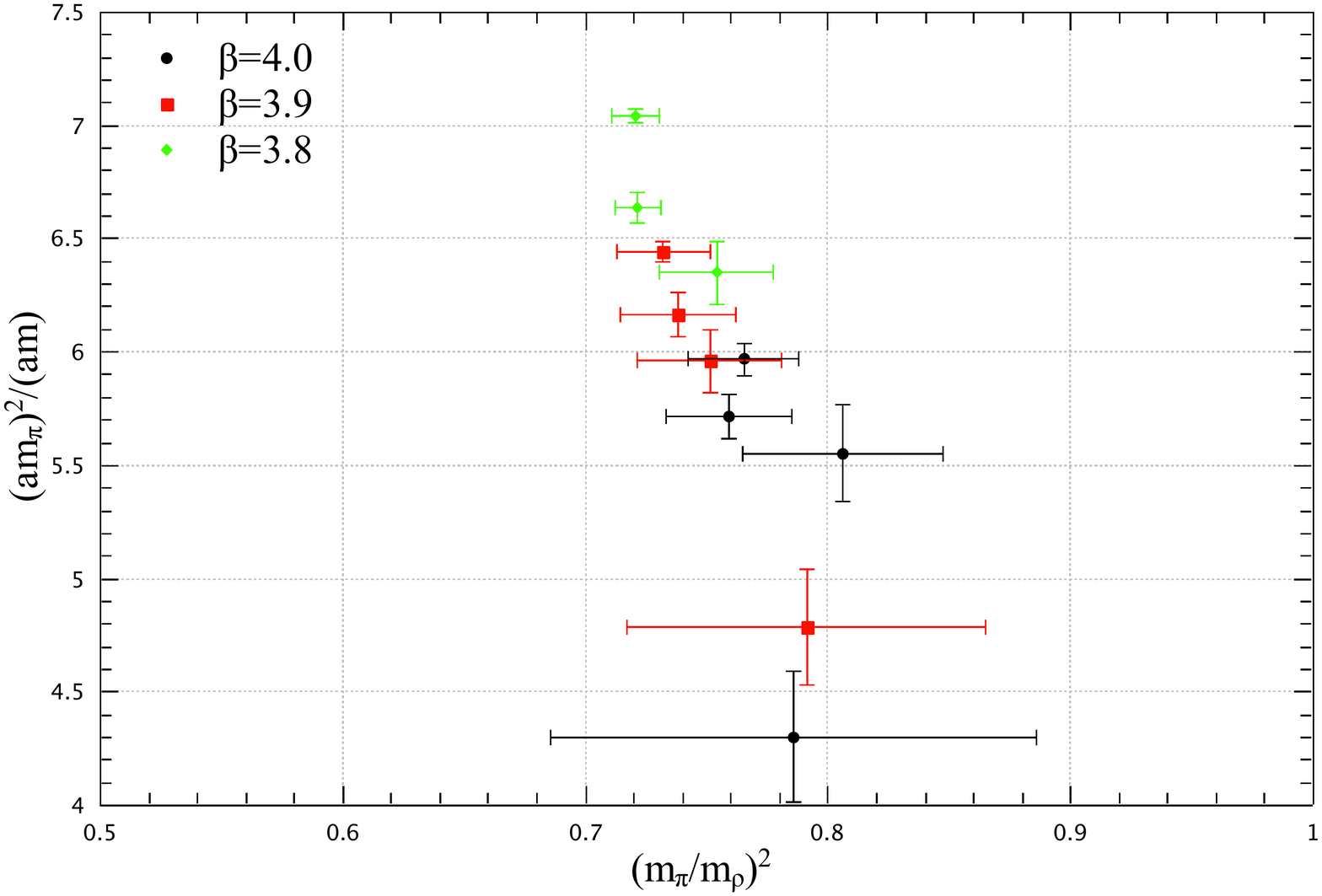}
\caption{Ratio $(am_\pi)^2/(am)$ as a function of $(m_\pi/m_\rho)^2$ for the $N_f=12$ data at $\beta_L=3.8$ (green diamonds), 3.9 (red squares) and $4.0$ (black circles).  }
\label{fig:CFTlike}
\end{center}
\end{minipage}
\end{figure}
A QCD scenario will draw a curve in this figure that extrapolates to the physical point for decreasing quark masses. What we observe in Fig.~\ref{fig:Edplot} suggests instead a behaviour that is to be expected for theories in the conformal window. The two mass ratios are ``stuck" at a tiny corner,  despite the fact that quark masses in the reported data vary on a rather wide range: $N_f=12$ bare masses from \cite{Deuzeman_PRD} and \cite{Fodor_Nf12} vary from $am=0.01$ to $am=0.07$ at various lattice couplings. $N_f=16$ bare masses from \cite{Damgaard:1997ut} vary from $am=0.025$ to $am=0.15$. 
All data in the Edinburgh plot are also away from the heavy quark limit and all have  $m_\pi/m_\rho\sim 0.8$, showing that all simulated masses are sufficiently light and fermions are dynamical.  
Fig.~\ref{fig:Edplot} also importantly suggests that all existing data for the $N_f=12$ spectrum  cover the same dynamical region; a comparison is therefore justified. 

Another interesting insight, useful to discriminate between a QCD-like and a conformal behaviour, can be gained through  Figs.~\ref{fig:QCDlike} and \ref{fig:CFTlike}.  At fixed lattice spacing, one can study the ratio $(am_\pi)^2/(am)$ as a function of $(m_\pi/m_\rho)^2$. This ratio  behaves as a constant in QCD to a good approximation, so that parallel horizontal lines are drawn at different lattice couplings. Fig.~\ref{fig:QCDlike} also shows that the ratio increases with decreasing $\beta_L$, a signal that the $\beta$ function for this theory is negative. 
Within the conformal window the behaviour should be quite different, and in fact analogous to what is observed in Fig.~\ref{fig:CFTlike}: the ratio should behave as $(am_\pi)^2/(am)\sim (am)^{2\delta}/(am)\sim (am)^{2\delta-1}$, with $0.5<\delta\lesssim 1$. 
Separate constant lines at different lattice spacings are thus no longer observed and data are concentrated around one value of $m_\pi/m_\rho$. Ideally, taken the same value of $m_\pi/m_\rho$ at two different lattice spacings, an ordering opposite to the QCD-like case would suggest a positive $\beta$ function. The latter is true at the strong coupling side of the IRFP. A few points in Fig.~\ref{fig:CFTlike} have a sufficiently close value of $m_\pi/m_\rho$  and show indeed the inverted ordering proper of a positive $\beta$ function. 
This is another way to look at the results for the $\beta$ function reported in \cite{Deuzeman_PRD}.
\begin{figure}[b]
\includegraphics[width=.90\textwidth]{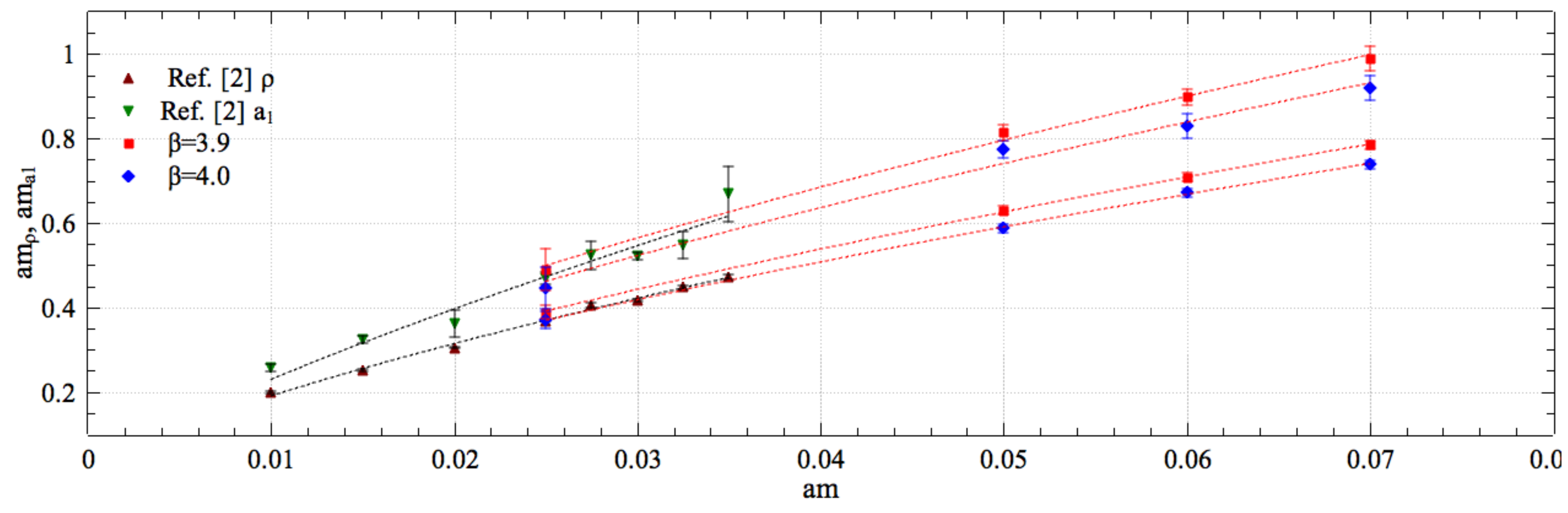}
\caption{Mass of the lightest axial-vector $m_{a_1}$ (top) and the lightest vector $m_\rho$ (bottom) as a function of the bare quark mass: $\beta_L= 3.9,\,4.0$ and $am=0.025,0.05,0.06,0.07$ (rightmost), $\beta_L=2.2$ and $am=0.01$ to $0.035$ from \cite{Fodor_Nf12} (leftmost).  Superimposed curves are fits to a power-law with zero intercept and free exponent.   } 
\label{fig:ma1mrho}
\end{figure}
\section{The vector and axial-vector mass splitting}
\label{sec:spectrum}
In \cite{Deuzeman_PRD} it was shown that the lightest pseudoscalar and vector masses at various simulated couplings were well fitted by a power-law with close exponents in the range 0.6-0.7, thus excluding the Goldstone nature of the pion and showing that data are away from the heavy quark regime.
As a word of caution, we add that the accuracies of the spectrum data and fits are  not  comparable, as of today, with  those  achieved  by  the  fits  to  the  chiral condensate in \cite{Deuzeman_PRD}. The latter have been shown to be at infinite volume within statistical uncertainties. For the spectrum, finite volume effects are expected to be present and of the order 
of about 10$\%$. 

The vector and  axial-vector mass splitting is another indicator of the restoration of chiral symmetry.  In Fig.~\ref{fig:ma1mrho} we show our data for $\beta_L=3.9$ and $4.0$, and  data from \cite{Fodor_Nf12} for the lightest vector $\rho$ and axial-vector $a_1$. The best fits with zero intercept and free exponent are  also reported. 
Best fit values of the exponents are $\delta_{a_1} = 0.67(4),\, \delta_\rho = 0.68(3)$,  at $\beta_L = 3.9$,  $\delta_{a_1} = 0.68(7),\, \delta_\rho = 0.67(3)$, at $\beta_L=  4.0$, and $\delta_{a_1} = 0.79(9),\, \delta_\rho = 0.72(2)$ for the data from \cite{Fodor_Nf12}. All exponents lie around 0.7.  A power-law fit with free intercept favours a slightly negative intercept with non unit exponent, thus disfavouring  a chirally broken scenario. 
The goodness of power-law fits with zero intercept for both vector and axial states suggests their degeneracy in the chiral limit, thus a restored chiral symmetry. 
Unfortunately, the $N_f=16$ results in \cite{Damgaard:1997ut} seem to be still affected by rather large finite volume effects, and for this reason we omit them here.
We are currently simulating $N_f=16$ with the same lattice action as $N_f=12$ \cite{Deuzeman_PRD} in order to meaningfully compare the two theories.
\section{Goldstone boson and chiral order parameter}
\label{sec:mpi_pbp}
\begin{figure}[h]
\includegraphics[width=.97\textwidth]{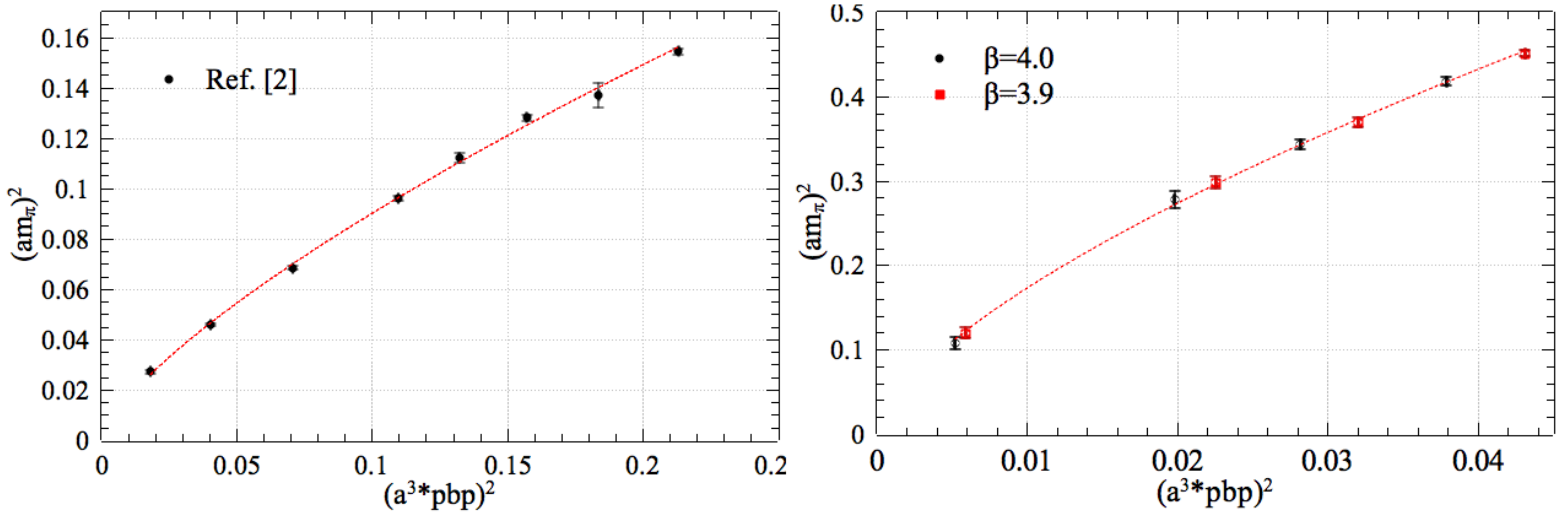}
\caption{$(a m_\pi )^2$ as a function of $(a^3 \langle\bar\psi\psi\rangle)^2$ for all existing $N_f=12$ data: (right) updated data from \cite{Deuzeman_PRD} for $\beta_L=3.9$ (red square)  and $\beta_L=4.0$ (black circle), (left) data from \cite{Fodor_Nf12} for $\beta_L=2.2$ (largest volumes only).   } 
\label{fig:Goldstone_pbp_All}
\end{figure}

It was observed in \cite{Deuzeman_PRD}  that an additional powerful discriminator between exact and spontaneously broken chiral symmetry is provided by the relation between the would-be Goldstone boson mass and the chiral condensate. The observation was based on the theoretical analysis of \cite{Kocic:1992is}. Here, we analyse all available $N_f=12$ data in light of these theoretical premises. 
\begin{figure}[t]
\begin{center}
\includegraphics[width=.45\textwidth]{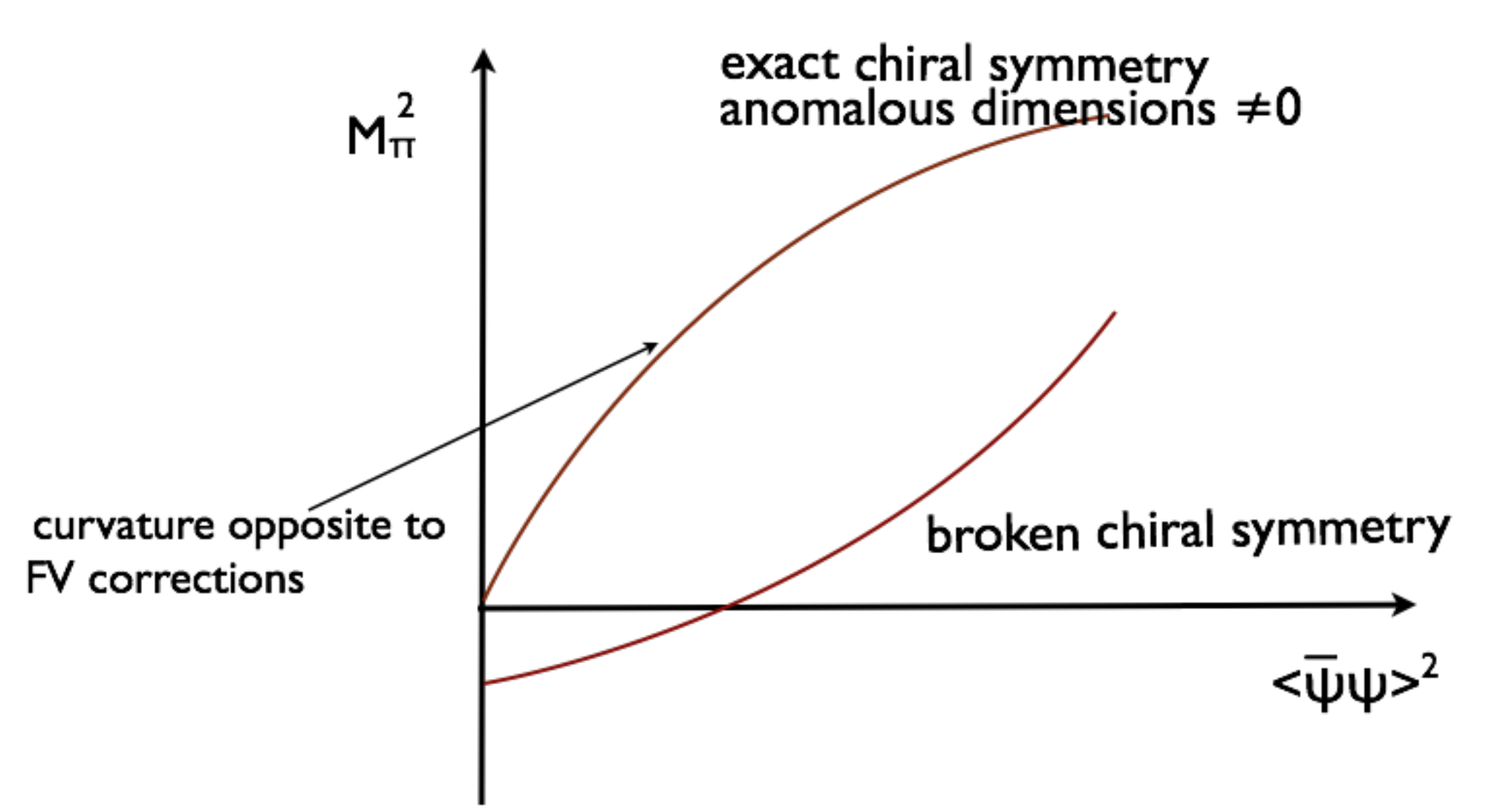}
\caption{The bare pseudoscalar mass squared as a function of the bare chiral condensate (order parameter) squared in a chirally symmetric (top) and chirally broken (bottom) phase \cite{Kocic:1992is}. } 
\label{fig:theory}
\end{center}
\end{figure}
In a chirally symmetric phase the behaviour of $m_\pi^2$ as a function of the chiral condensate is as illustrated in Fig.~\ref{fig:theory} (upper curve): it has positive curvature, it extrapolates to zero, and the curvature is due to non zero anomalous dimensions. Importantly, the curvature is opposite to that induced by finite volume effects. Hence, it provides at the same time a powerful indicator of the presence of finite volume corrections for lattice data. 
In a chirally broken phase the curvature is opposite and it extrapolates to a negative value, 
Fig.~\ref{fig:theory} (lower curve).  In Fig.~\ref{fig:Goldstone_pbp_All}  we collect 
our data (right) and data from \cite{Fodor_Nf12} (left).
Both data sets clearly show a positive curvature, and best fits to a power-law  $(a m_\pi)^2 = A (a^3 \langle\bar\psi\psi\rangle)^{2\delta_\chi}$ with zero intercept do not suggest qualitative differences between data sets.
We obtain a best fit exponent $\delta_\chi =0.66(2)$ for joint data sets at $\beta_L =3.9$ and $4.0$ (see also \cite{Deuzeman_PRD}) and  $\delta_\chi =0.727(5)$ for the data from \cite{Fodor_Nf12}.
\section{Conclusions}
\label{sec:conclusions}
We have reported on the spectrum of the $SU(3)$ gauge theory with twelve flavours in the fundamental representation. In particular, we have isolated a few signatures that are useful to discriminate between the hadronic phase, i.e. QCD at zero temperature, and the conformal phase - proper of theories within the conformal window. In this analysis we have assumed that all our lattice data, even if inside the conformal window, are away from the IRFP and correlators follow a leading exponential decay law; this seems to be supported by the obtained results. Further investigations at weaker coupling, lighter masses (and longer temporal extents) would provide a useful piece of additional information. We have analysed the ratio $m_\pi/m_\rho$, the Edinburgh plot and the $\rho-a_1$ mass splitting for all existing data. Points in the Edinburgh plot stick to a region $m_\pi/m_\rho \sim 0.8$, despite covering a wide range of bare lattice fermion masses. Fits to the $\rho -a_1$ splitting favour the degeneracy of the two states in the chiral limit. Finally, we have re-proposed the relation between the Goldstone boson mass and the chiral order parameter as a powerful indicator of restored  chiral symmetry. All existing data 
for the $N_f=12$ theory seem to consistently favour chiral symmetry restoration and an almost universal power-law behaviour for all massive states, which is to be expected inside the conformal window.
\section*{Acknowledgements}
This work was in part based on the MILC collaboration's public lattice gauge theory code. Simulations were performed on the IBM BG/P at the University of Groningen and the IBM Power6+ Huygens at SARA (Amsterdam), under support of the Dutch NCF.

\end{document}